\documentclass[twocolumn]{aa}
\usepackage{graphicx}
\begin{document}
%
%
%
   \title{Phase diversity restoration of sunspot images}

   \subtitle{I. Relations between penumbral and photospheric features}

   \author{J.A. Bonet       \inst{1}
          \and
          I. M\'arquez      \inst{1,4}
          \and
          R. Muller         \inst{2}
          \and
          M. Sobotka        \inst{3}
          \and
          A. Tritschler     \inst{5,6}
          }

   \offprints{J.A. Bonet}

   \institute{Instituto de Astrof\'\i sica de Canarias, 38205 La Laguna,
                Tenerife, Spain
         \and
             Observatoire du Pic du Midi, 57 Avenue d'Azereix, B.P.826,
                65008 Tarbes, France
         \and
             Astronomical Institute, Academy of Sciences of the Czech Republic,
                CZ-25165 Ond\v rejov, Czech Republic
         \and
             Departamento de An\'alisis Matem\'atico de la Universidad
             de La Laguna, E-38271 La Laguna, Tenerife, Spain
         \and
             Kiepenheuer--Institut f\"ur Sonnenphysik, Sch\"oneckstr. 6, 
             79104 Freiburg, Germany
         \and
             Big bear Solar Observatory, New Jersey Institute of Technology, 
             40386 North Shore Lane, Big Bear City, CA-92314, U.S.
	}

   \date{Received 20 October 2003 / Accepted 30 April 2004}

   \abstract{
   We investigate the dynamics of and the relations 
   between small-scale penumbral and photospheric features near the outer
   penumbral boundary: penumbral grains (PGs), dark penumbral fibrils,
   granules, and photospheric $G-$band bright points. The analysis is based
   on a 2\,h time sequence of a sunspot close to disc center,
   taken simultaneously in the $G-$band and in the blue continuum at
   450.7\,nm. Observations were performed at the Swedish 
   Vacuum Solar Telescope (La Palma) in July 1999. A total of 2564 images
   (46\arcsec $\times$ 75\arcsec) were corrected for telescope aberrations and
   turbulence perturbations by applying the inversion method of phase diversity.
   Our findings can by summarized as follows: 
   (a) One third of the outward-moving PGs pass through the outer penumbral
   boundary and then either continue moving as small bright features or 
   expand and develop into granules. (b) Former PGs and $G-$band bright
   points next to the spot reveal a different nature.
   The latter have not been identified as a continuation of PGs escaping 
   from the penumbra. The $G-$band bright points
   are mostly born close to dark penumbral fibrils where the magnetic field is
   strong, whereas PGs stem from the less-magnetized penumbral component and
   evolve presumably to non-magnetic granules or small bright features.

   }
   \titlerunning{Phase diversity restoration: Penumbral and photospheric
                features}
   \maketitle
%

\section{Introduction}

The magnetic field generated at the bottom of the convection zone
emerges in active regions, from where it is diffused to the surface
of the Sun, by convective motions and large scale plasma circulation.
However, the mechanism by which magnetic flux is taken away from sunspots
is not well known yet. The aim of this work is to investigate how the
features inside and outside the penumbra of a decaying sunspot are 
related. Decaying sunspots are well suited for that purpose, because it
is believed that the decay of a sunspot proceeds predominantly 
by erosion of its perimeter leading to a quadratic decay law 
(Sheeley 1972; Meyer et al.\ 1974; Petrovay \& Moreno-Insertis 1997).

Sunspots are very complex magnetic structures, embedded in a convective
plasma, which show a small-scale pattern, down to the resolution limit
of the largest contemporary solar telescopes ($\sim$0\farcs1 = 75 km). The
surrounding convective plasma is mainly observed in the form of granules.
However, the granular convection is perturbed by the presence of many
magnetic elements, visible in $G-$band as tiny bright points located
in intergranular spaces. They correspond to the facular and magnetic
elements which are visible in lower resolution filtergrams and
magnetograms. The granules (Muller \& M\'ena 1987; Simon et al.\ 1988) 
and the magnetic elements -- either
observed in filtergrams or in magnetograms -- (Sheeley 1969, 1972;
Sheeley \& Bhatnagar 1971; Vrabec 1971, 1974; Harvey \& Harvey 1973;
Muller \& M\'ena 1987; Brickhouse \& La Bonte 1988; Lee 1992;
Ryutova et al.\ 1998 ) move radially outwards, through an annular cell.
This annular cell has been called a ``moat" and the magnetic elements,
moving accross, ``moving magnetic features" or ``MMFs" ( Sheeley 1972;
Harvey \& Harvey 1973).
Dopplergrams have shown that this outflow, with speed of 0.5--1.0
km~s$^{-1}$, is very similar to that observed in supergranules.
This has suggested that the moat is presumably a supergranule whose
center is occupied by a sunspot, and which extends approximately
10000 km beyond the spot's edge (Sheeley 1972). The penumbra--granulation
boundary appears very sharp, since the photospheric features
(granules and bright points) are directly adjacent to darker penumbral
elements (dark diffuse patches, dark and bright fibrils).

The penumbra has a very complex morphological, magnetic and dynamic
structure, which is not yet properly understood physically. Visually,
with the spatial resolution of the present work ($\sim$0\farcs23), it
appears to be formed by narrow, nearly radial bright filaments, separated
by dark fibrils. Local brightenings, called penumbral grains (PGs) are
observed in the bright filaments. In images of extremely high
resolution (better than 0\farcs12) obtained with the 1-m Swedish Solar
Telescope at La Palma, Scharmer et al.\ (2002) detected dark
cores inside the bright filaments.

The magnetic field in the penumbra is as strong as 1500 G at the
inner boundary, and decreases to 700 G at the outer boundary, becoming 
more and more horizontal. In addition, it is very inhomogeneous
in the azimuthal direction. From the synthesis of many papers, sometimes presenting conflicting results, 
it appears that, in the outer penumbra, higher field strengths tend to be 
correlated with more horizontal fields (by 10 to 20 degrees) and dark intensity
structures (see e.g. Beckers \& Schr\"oter 1969; 
Westendorp Plaza et al.\ 2001; Bellot Rubio 2003 and references therein).
In addition to that, several authors report that the magnetic radius of
sunspots extends beyond the visible boundary of the penumbra (Skumanich
1992; Title et al. 1993; Mart\'\i nez Pillet 1997; Westendorp Plaza et al.\
2001; Bellot Rubio 2003).

The penumbra of sunspots is not only a magnetic, but also a very
dynamic plasma: The bright PGs and dark features both move, either
toward or away from the umbra; an irregular and not steady outflow of
gas (the Evershed effect) is detected. In the inner penumbra, the PGs
always move towards the umbra, and reach a maximum velocity of
0.5 km s$^{-1}$ at the umbra-penumbra boundary (Muller 1973).
In the outer penumbra, they have been found to move
either towards the surrounding photosphere (Wang and Zirin 1992;
Denker 1998; Sobotka et al.\ 1999; Sobotka \& S\"utterlin 2001),
or towards the umbra (Muller 1973; T\"onjes \& W\"ohl 1982; Zirin
\& Wang 1989). The dividing line, separating the two opposite motions, 
is located near the outer boundary at about 1/3 of the
width of the penumbra.
Schlichenmaier (2002) elaborated a moving flux tube model developing
waves and kinks (a photospheric ``serpent") that reproduces both
directions of motions of the PGs.

The present paper is the first in a series based on two excellent 2-hour
time series of sunspot images reconstructed with the phase-diversity
technique. The long duration of these series, together with the high
and stable quality throughout the whole period, which is substantially
improved after the reconstruction (see Fig.~1), makes this material
one of the best data sets ever produced to study the morphology and the dynamical
behaviour of a sunspots fine structure and its surroundings.
In addition, the strict simultaneity of the images in the $G-$band and
the blue continuum significantly increases the reliability of the identification and
tracking of $G-$band bright points. 

In this work we are interested in the relation between the
penumbral and the photospheric features as they move across the boundary
of a decaying sunspot. Such a decaying sunspot is well suited for our
purpose, because it has a simple, nearly symmetric configuration,
surrounded by a well developed and nearly circular moat.
Moreover, the features in the outer penumbra and in the
moat move in the same direction, outward, which is a favourable situation
to find out if there is a continuity in their properties. In the second paper 
of the series, we will investigate in detail the motions of granules,
families of granules formed by recurrently splitting granules, and
small-scale magnetic bright points, in order to see how they behave and
interact when they are dragged by a supergranular convective flow.

\section{Observations and Data Processing}

\subsection{Observations}

The leading sunspot of a decaying bipolar group NOAA 8620 was
observed on 7 July 1999 at the Swedish Vacuum Solar Telescope,
La Palma (Scharmer et al.\ 1985), near the center of the
solar disc at $\mu = 0.93$. The sunspot showed a regular shape and a slow and 
smooth area decay from July 3 to July 13 as recorded in the daily area 
values of Solar Geophysical Data.

Two strictly simultaneous time sequences of high-resolution images
were taken in the wavelength bands at
$450.7 \pm 0.45$ nm  (blue) and $430.8 \pm 0.55$ nm  ($G-$band).
The observing sequence spans for more than two hours (1282 images
per series), from 7:30\,UT to 9:43\,UT. The images were acquired using
two Kodak Megaplus 1.6 CCD cameras, with a dynamical
range of 10 bit and $1536 \times 1032$\,pixels. The pixel size and
exposure time were 0\farcs 083 and  $\sim$40\,ms, respectively. The 
cameras were placed in orthogonal light paths produced by a cubic beam 
splitter. A common mechanical shutter in front of the beam splitter ensured
the simultaneous exposure in both cameras.

To apply the phase-diversity technique for image reconstruction, pairs of 
focused and intentionally defocused images must be taken
simultaneously. To that 
aim each CCD camera is equipped with a special beam-splitter that images 
the focused-defocused image pair onto the same chip (see Fig.~2 in L\"ofdahl
et al. 1998). The defocus is caused by a slightly longer optical path length in 
one of the split beams. In the present case, the equivalent
defocusing displacement in vacuum was 8.85\,mm  that corresponds to a phase 
shift at the edge of the aperture of 1.15\,waves and 1.10\,waves at
430.8\,nm and 450.7\,nm, respectively. After selection of the common
area in both images of the pair, the removal of spurious effects
induced by the beam splitter, and by the apodization during restoration 
process at the edges of the image, the effective field-of-view 
results in 46\arcsec $\times$ 75\arcsec \ (560 $\times$ 910\,pixels).
The cameras were working in a real-time frame-selection mode storing the best
4 images within a time interval of 18\,s. Considering the additional time to
transfer the data to the hard disc, the resulting mean time interval between
two consecutive sets of 4 images was 25\,s.

\begin{figure*}
\centering
\includegraphics[width=14cm]{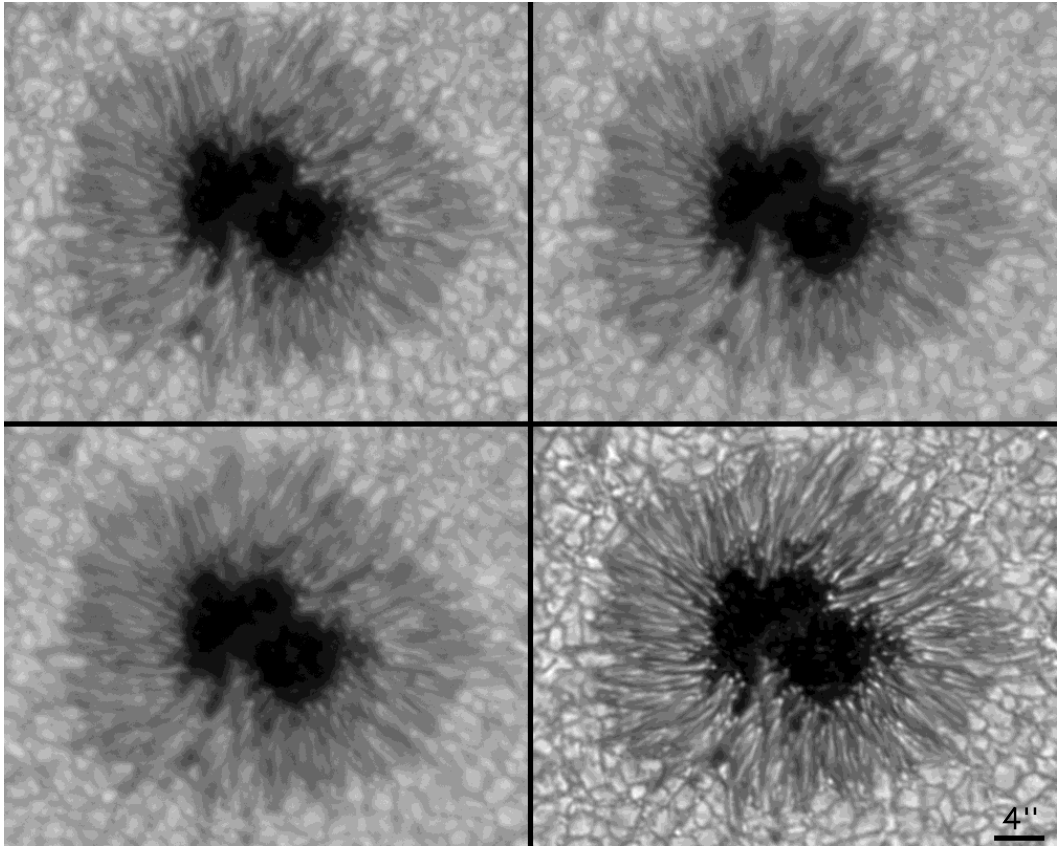}
\caption{Sunspot images taken from the sequence at 450.7\,nm.
 Lower right panel: Central part of the restored image resulting from 
 the combination of three raw images observed close in 
 time and shown in the other three panels of the figure. Each 
 raw image has been individually inverted with phase diversity 
 in order to evaluate its corresponding instantaneous OTF.}
\end{figure*}

In an independent third channel, a complementary data set was acquired
in the Ca II K line ($\lambda \, 393.3 \pm 0.15$ nm ) for monitoring the
magnetic activity in the observed field.
Standard flat-field and dark-current corrections were applied to all 
images before the image restoration.

\subsection{Data Processing}

As mentioned above, the phase-diversity (PD) technique has been applied to 
correct the images of the blue and $G-$band sequence up to close the 
diffraction limit. This technique allows one to estimate both the unknown 
observed object and the wavefront aberration.
A few papers in the literature are complementary in the description of the
PD technique for image reconstruction (e.g. Gonsalves \& Childlaw
1979; Gonsalves 1982; Paxman, Schulz \& Fienup 1992; 
L\"ofdahl \& Scharmer 1994: and Paxman et al. 1996).
However, we summarize the most relevant items of the 
method as an introduction to this powerful tool for high resolution 
solar imaging and to report on our particular setting of parameters and 
assumptions to run the computing code.

This technique requires the detection of at least two images of the 
object. One of these is the conventional focal-plane image that is degraded 
by unknown aberrations induced by the atmosphere and the telescope. 
The other one is a simultaneous image affected by the same unknown 
aberrations plus an extra known aberration (the easiest controlled aberration 
to produce is a defocus). Both image realizations, 
$i_{1}(\mbox{\boldmath $q$})$ and $i_{2}(\mbox{\boldmath $q$})$, can be 
mathematically represented by the system of equations

\begin{equation}
\left. \begin{array}{c}
   i_{1}(\mbox{\boldmath $q$}) = i_{0}(\mbox{\boldmath $q$})\ast 
s_{1}(\mbox{\boldmath $q$}) + n_{1}(\mbox{\boldmath $q$}), \\ 
   i_{2}(\mbox{\boldmath $q$}) = i_{0}(\mbox{\boldmath $q$})\ast 
s_{2}(\mbox{\boldmath $q$}) + n_{2}(\mbox{\boldmath $q$}), \\
   \end{array} \right\}
\label{eq:div1}
\end{equation}

where $\ast$ stands for convolution and $\mbox{\boldmath $q$}$ is the vectorial
notation for the coordinates of the image points; $i_{0}$ is the ``true'' object
and $i_{1}$, $i_{2}$ are the two observed images; $s_{1}$
and $s_{2}$ represent the corresponding point spread functions (PSF) of the
optical transmission system -- including the
terrestrial atmosphere and the telescope -- and $n_{1}$, $n_{2}$ are additive
terms of noise.

Gonsalves \& Childlaw (1979) propose the solution of this system of equations 
as a least-squares fit which in the Fourier domain can be written as the 
following error metric:

\begin{equation}
\left. \begin{array}{c}
   L(I_{0},{S}_{1}) = \sum\limits_{\mbox{\boldmath $u$}} \Bigl\{ \big\vert
   I_{1}(\mbox{\boldmath $u$}) - {I}_{0}(\mbox{\boldmath $u$}){S}_{1}
   (\mbox{\boldmath $u$}) \big\vert ^{2}  \\ 
   ~~~~~~~~~~~~~~~~~~~~~~~~~ + \gamma \, \big\vert I_{2}
   (\mbox{\boldmath $u$}) - {I}_{0}(\mbox{\boldmath $u$}) {S}_{2}
   (\mbox{\boldmath $u$}) \big\vert ^2 \Bigr\},  \\
\end{array} \right.
\label{eq:div3}
\end{equation}

where capital letters represent the Fourier transform of the corresponding
functions denoted by lower-case characters and ${\mbox{\boldmath $u$}}$ is
the frequency vector in the Fourier domain. $\gamma $ is included as a factor 
to equalize the noise in both images (a ratio of the  respective noise
variances). Typical values for noise level in our data at both wavelengths 
are in the range 0.6--0.8\,\% of the average intensity. $S$ represents the 
optical transfer function (OTF). It is directly related to the phase 
aberration in the pupil (see e.g. Bonet 1999), which in turn can be 
approximated by the first $J$ terms of a Zernike expansion. In the present case
 Zernike modes 2-21 have been employed in the expansion, i.e. 20 coefficients, 
$\mbox{\boldmath $\alpha$} \equiv \left\{ \alpha_{j}, j=2,...,21\right\}$ 
have to be determined to estimate the wave front error and subsequently 
$S(\mbox{\boldmath $u$},\mbox{\boldmath $\alpha$})$. Using a greater number 
of modes, e.g. $J=45$, did not significantly improve the results while it 
substantially increased the computing time.

Part of the minimization of (\ref{eq:div3}) can be performed analytically. 
The solution of equation: $\partial L / \partial {I}_{0} = 0$, 
gives an estimate of the object, $\hat{I}_{0}$, that minimizes
(\ref{eq:div3}) for fixed \mbox{\boldmath  $\alpha$}.
\begin{equation}
   \hat{I}_{0}(\mbox{\boldmath $u$}) = {I_{1}(\mbox{\boldmath $u$})\,
   {S}_{1}^{*}(\mbox{\boldmath $u$},\mbox{\boldmath $\alpha$}) + \gamma \,
   I_{2}(\mbox{\boldmath $u$})\,{S}_{2}^{*}(\mbox{\boldmath $u$},
   \mbox{\boldmath $\alpha$}) \over \vert {S}_{1}(\mbox{\boldmath $u$},
   \mbox{\boldmath $\alpha$}) \vert ^2 + \gamma \,
   \vert {S}_{2}(\mbox{\boldmath $u$},\mbox{\boldmath $\alpha$}) \vert ^2}.
\label{eq:div4}
\end{equation}
The subsequent substitution in (\ref{eq:div3}) leads to the modified error
metric $L_M (\mbox{\boldmath $\alpha$})$
\begin{equation}
   L_{M}(\mbox{\boldmath $\alpha$}) = \sum_{\mbox{\boldmath $u$}}
   {\big\vert I_{1}(\mbox{\boldmath $u$})\,
   {S}_{2}(\mbox{\boldmath $u$},
   \mbox{\boldmath $\alpha$}) - I_{2}(\mbox{\boldmath $u$})\,
   {S}_{1}(\mbox{\boldmath $u$},\mbox{\boldmath $\alpha$})\big\vert^2
   \over \vert {S}_{1}(\mbox{\boldmath $u$},\mbox{\boldmath $\alpha$})
   \vert ^2 + \gamma \, \vert {S}_{2}(\mbox{\boldmath $u$},
   \mbox{\boldmath $\alpha$}) \vert ^2},
\label{eq:div5}
\end{equation}
which is now independent of the object Fourier transform ${I}_{0}$.
Thus, the parameter space over which the optimization is performed has
dimension 20. Once these parameters are determined, 
$\hat{S}_{1}$ and $\hat{S}_{2}$ can be constructed and the object estimate, 
$\hat{I}_{0}$, can be derived from (\ref{eq:div4}), thereby completing the 
image reconstruction process.

Poor seeing may produce a low signal-to-noise ratio and very small
and uncertain values of $S(\mbox{\boldmath $u$})$ at certain spatial
frequencies. At these frequencies, Eq.~(\ref{eq:div4}) produces 
an excessive amplification of the signal, giving rise to artifacts that 
show up as a regular pattern in the restored scene. To circumvent this 
problem, we combine in a sort of speckle summation 
(see e.g. Paxman et al. 1996) the results of the inversion of 3 or 4 
image pairs taken close in time, so that evolutionary aspects in the 
solar structures can be neglected. The lower right panel of Fig.~1 shows 
the central part (41\farcs5 $\times$ 33\farcs2) of the final restored image 
resulting from the combination of three single-image inversions of the blue 
sequence. These images were recorded within a period of 18\,s.

The restored images were de-rotated to compensate for rotation of
the field of view in the focal plane induced by the alt-azimuthal configuration
of the telescope. Furthermore, the images were aligned, destretched 
(computer code by Molowny-Horas \& Yi, 1994), and filtered for
p-modes in the $k-\omega$ space (cut-off phase velocity 5\,km\,s$^{-1}$).
Two movies were produced, for the blue continuum and the $G-$band,
of 288 frames each, spanning over 2 hours ($\Delta t = 25$\,s) and covering
a field of 46\arcsec $\times$ 75\arcsec , with the
sunspot located in the center. Direct inspection of the images
reveals structures up to about the Rayleigh resolution limit at both
working wavelengths ($\sim$0\farcs23).

A feature-tracking technique (Sobotka et al.\ 1997) was used to
measure positions, intensities and sizes of PGs and of $G-$band
bright points. PGs were tracked in the series of blue images.
To isolate them from other structures, photospheric granules
and umbral dots were masked out and a segmentation algorithm
was applied to each blue image, based on the rule that pixels
with downward concavity, representing bright features, are set
to 1 and the rest to 0 (cf.\ Sobotka \& S\"utterlin 2001).
The obtained binary masks were then multiplied by corresponding
blue images, so that original intensities of PGs were preserved.

To isolate $G-$band bright points, the destretch algorithm was
applied to each blue image to align it exactly with the structures
in the corresponding $G-$band frame. The granular contrast of
aligned blue images was reduced by a factor of 0.833 to match the
contrast observed in the $G-$band frames. Then the blue
frames were subtracted from the $G-$band frames, thus producing
intensity difference images with the $G-$band bright points
substantially enhanced. The segmentation was done by thresholding
these difference images, obtaining binary masks, and by multiplying
the masks by the original $G-$band frames.

After the segmentation, the features under study (PGs and
$G-$band bright points) are formed from the non-zero intensity
pixels. In the next step, pixels forming a feature are labelled
by an identification number. Then, the spatial coincidences of
features in each pair of subsequent images are investigated.
Two features are identified as predecessor/successor if they
coincide in the coordinates of at least one pixel in both frames.
The maximum intensity, its coordinates and the total number
of pixels in the feature are recorded for each frame.
The lifetimes (number of frames) and the sizes (number of pixels)
of the features can be obtained directly from the records.
Formation, death, splitting and merging of features are taken
into account. In the case of splitting, the brightest feature
is adopted as the successor, while if merging occurs, the merged
feature is defined to be the continuation of the feature with
the longest record.

To improve the reliability of the tracking results and to avoid
small-scale noise in the segmented images, for the further analysis
we select only features with lifetimes and time-averaged sizes above
certain minimum values (see Sect.\ 3), where the minimum size is
usually derived from the resolution limit of the observations.
Variable image quality may cause spurious merging and splitting
that can make independent features appear related. This would
lead to large displacements of the maximum intensity positions
and to unrealistic velocities of proper motions. This problem
can be partially fixed by selection of features with time-averaged
velocities smaller than the cut-off velocity used
in the $k-\omega$ filter. Time-averaged velocities are calculated
using linear least-squares fits to the positions. More additional
criteria can be applied to select features with specific
characteristics.

Finally, a ``visual consistency check'' is applied to the selected
features: Trajectories are checked visually to eliminate discontinuities
in position or strong bends/breaks in the trajectory, caused by spurious
coincidences of different features, and to discard such features
that are not the subject of the study.

\section{Results}

\subsection{Outward-moving penumbral grains}

We have already mentioned in Sect.\ 1 that in the outer 1/3 of the
penumbra most of the PGs move towards the outer penumbral border
(P/G boundary).

A natural question is what happens when they reach the boundary between 
the penumbra and the surrounding photospheric granulation (P/G boundary).
A visual inspection of a movie composed of the blue images shows that
the outward-moving PGs evolve in three possible ways:\\
1. They disappear in the penumbra before reaching the P/G boundary.\\
2. They cross the P/G boundary and continue to move in the granulation
as small bright features.\\
3. After crossing the P/G boundary, they increase their size
and develop into granules that move away from the sunspot.\\

This different behaviour of the PGs crossing the P/G boundary is quite 
remarkable and justifies the separate study of the features that are
observed to expand and those that do not expand.

To study the evolution of PGs quantitatively, the mentioned above 
feature-tracking technique was applied to a field of 12\farcs5 $\times$ 20\farcs8,
including a part of the penumbra and adjacent granulation. This region
is located on the left-hand side of the sunspot in Fig.~1 (see also
Fig.~2 showing the contour of the time-averaged outer penumbral
border in this region). The criteria to select bright features of interest 
from the raw results of the tracking (see Sect.\ 2) were as follows:
(a) Minimum lifetime of 250\,s, making it possible
to pass a distance of 1000\,km with the (b) maximum allowed average
speed of 4\,km\,s$^{-1}$; (c) minimum average size of 0\farcs28,
corresponding to an area of 9\,pixels. To eliminate from the tracking
results most of the granules and PGs that probably do not cross
the P/G boundary, an additional criterion (d) of average brightness
ranging from $I=0.9$ to 1.15 (in units of the mean photospheric
intensity) was applied.
This criterion was derived from the histogram of time-averaged
intensities (resulting from the tracking), which clearly showed two
populations, PGs and granules, with peaks at $I=0.92$ and 1.16,
respectively, and with a dip at $I=1.02$. We can expect that
the time-averaged intensities of PGs that cross the P/G border
and penetrate into the granulation correspond to the overlap
of both populations. This way, 712 bright features were
selected. Finally, the visual consistency check was applied
to these features and all inward moving PGs, together with features
originating outside the penumbra were discarded.

After the visual consistency check we obtained a sample of 126
outward moving PGs, including their trajectories. The noise in the
measured positions was removed by smoothing the trajectories
using cubic splines. Instantaneous velocities were calculated
as derivatives of the smoothed trajectories. Each PG was inspected
individually to check its positions in time with respect to
the P/G boundary. About 2/3 of PGs disappear before reaching the P/G
boundary, 1/6 cross the boundary and continue moving as small bright
features with a diameter less than 0\farcs5, and about 1/6 cross the
boundary, expand in size, and develop into granules --here we use the term 
granules because the visual inspection of the movie reveals that these 
structures, when present in the photosphere, do behave as normal granules.
All the features that cross the P/G boundary preserve their outward
motion. Their trajectories, smoothed by cubic splines, are plotted
in Fig.~2 together with the time-averaged contour of the P/G border. 
This contour is shown only for illustration, because the ragged border
of the penumbra evolves in time and only individual tracking of PGs
gives the information if they crossed the P/G boundary or not.
The origins of trajectories are marked by asterisks for small features
and by squares for granules. Dots denote the following positions;
in the case of slow motions the dots merge, forming solid lines.
In general, the birth places of crossing and non-crossing PGs
are mixed. However, on average, crossing PGs in our sample
originate at the distance of about 2\arcsec \, from the P/G boundary
while the non-crossing ones by 0\farcs9 deeper in the penumbra
(the width of the penumbra is approximately 8\arcsec--10\arcsec).
The present sample of PGs does not include all PGs in the field of view.

The total lifetimes (including the penumbral and photospheric stages)
of the PGs that convert into small bright features are in the range
of 4.6--41.0\,min with the mean of 13.9\,min. The total lifetimes
of the PGs that develop into granules are longer: 5.9--120.0\,min,
with an average of 36.0\,min. For comparison, the non-crossing PGs live
4.2--102.5\,min, with a mean lifetime of 14.3\,min.
After crossing the penumbral border, small features have lifetimes
in the range 2.5--22.1\,min with the mean of 7.7\,min.
For granules which arise from PGs, the range is 1.7--36.3\,min
and the mean is 15.0\,min.
Alissandrakis et al. (1987) give a mean lifetime of photospheric
granules of 16\,min. Hence, after crossing the border, granules
of penumbral origin have a mean lifetime comparable with granules
formed in the quiet photosphere. Small features live shorter than
``average'' granules but longer than small ($<$ 0\farcs5) ones,
with a lifetime that was estimated at 1--3\,min (Kawaguchi 1980;
Hirzberger et al.\ 1999). 

The lifetime-averaged speeds of small bright features range from
0.4 to 2.7\,km\,s$^{-1}$ with a mean of 1.4\,km\,s$^{-1}$. The newly
formed granules are slower: 0.2--1.9\,km\,s$^{-1}$, with a mean speed 
of 1.1\,km\,s$^{-1}$. These velocities are slightly higher
compared to the average speed of all outward moving
PGs near the P/G border (0.9\,km\,s$^{-1}$, Sobotka \& S\"utterlin
2001) and to the horizontal speed of granules close to sunspots
(0.5--0.8 \,km\,s$^{-1}$, Muller \& M\'ena 1987, Simon et al.\ 1988).

Figure 3 demonstrates the transformation of an
outward moving PG into a small bright photospheric feature ({\it
top}) and into a granule ({\it bottom}). Dark dots in the 
figure denote intensity maxima of the features.

\begin{figure}
\centering
\includegraphics[width=8cm]{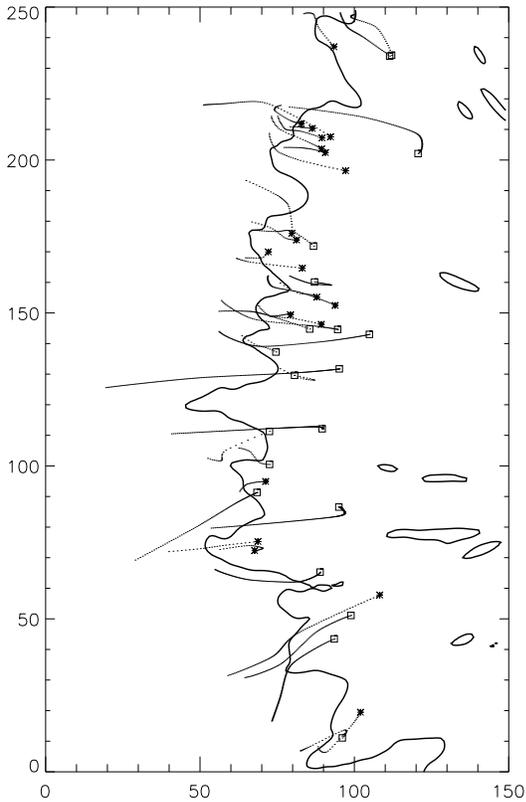}
\caption{Trajectories of PGs that crossed the P/G boundary,
 approximated in the figure by a time-averaged contour.
 Origins of trajectories are marked by asterisks for small features
 and by squares for granules. The coordinate unit is 1 pixel, i.e.,
 0\farcs 083.}
\end{figure}

\begin{figure*}
\centering
\includegraphics[width=14cm]{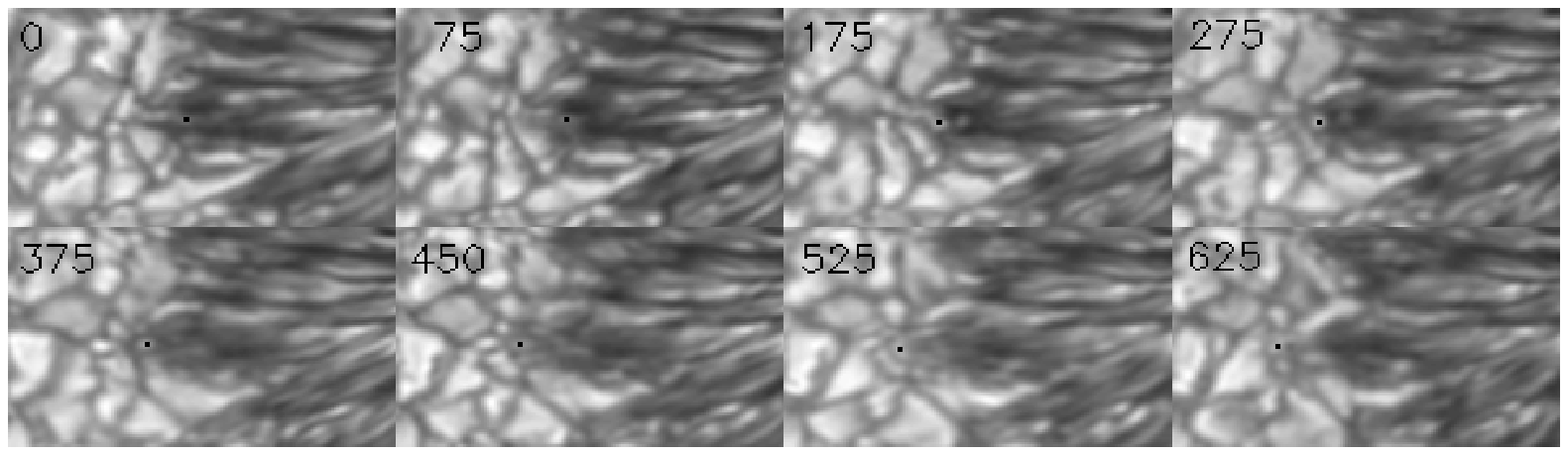}  
\vspace{1mm}
\includegraphics[width=14cm]{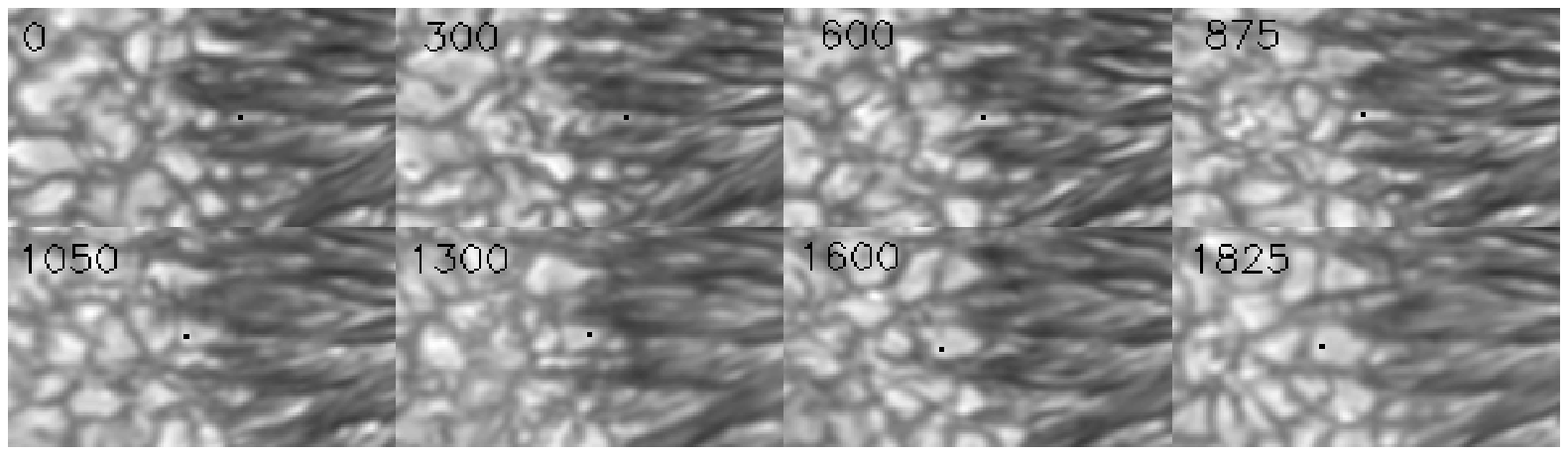}  
\caption{Series of frames showing two examples of PGs crossing the
 P/G boundary and converting into a small bright feature (top) and
 into a granule (bottom). The tracked features are marked by small
 black dots at the positions of intensity maxima. Numbers in the upper
 left corners show the elapsed time
 in seconds. The field of view is 12\farcs 5 $\times$ 7\farcs 5.}
\end{figure*}

In Table 1 we summarize the numbers of PGs that disappear in the penumbra
before crossing the P/G boundary, PGs that cross the P/G border and
convert into small bright photospheric features, and those that develop
into granules. We also show the numbers of PGs that accelerate
or decelerate their motion before and during the crossing of the
boundary. Note that most of the PGs which convert into
small features decelerate before they escape from the penumbra,
while those that expand and develop into granules accelerate.

\begin{table}
\caption{Summary of PGs  near the P/G border}
\begin{tabular}{lrrr}
\hline
\noalign{\smallskip}
PGs &  Total number &  Accel.  &  Decel.  \\
\noalign{\smallskip}
\hline
\noalign{\smallskip}
Non-crossing & 85 (67\,\%) &      &     \\
$\rightarrow$ small features & 21 (17\,\%) &   6  &  13 \\
$\rightarrow$ granules       & 20 (16\,\%) &  11  &   6 \\
\noalign{\smallskip}
\hline
\end{tabular}
\end{table}

\subsection{Relation of $G-$band bright points to the penumbra}

Numerous $G-$band bright points were observed in the sunspot moat,
which, in our case, had approximately an annular shape and an
average width of 8\arcsec. This has been determined from the 
horizontal velocity field of granules around the spot. The details 
will be described in the second paper of this series.
The feature tracking algorithm was applied
to the series of segmented $G-$band frames, detecting 776 $G-$band
bright points with lifetimes longer than 2.5 min and with time-averaged
sizes larger than 0\farcs28.Most of the $G-$band bright points (95\,\%) 
move away from the sunspot. A detailed analysis of their trajectories 
and horizontal velocities will be published in the second paper of 
this series.

To find a possible relation of the $G-$band bright points with the
penumbra and penumbral features, we looked for such $G-$band bright
points that were born in the neighbourhood of the penumbral border.
The neighbourhood was defined by the maximum distance of 0\farcs 25
(3 pixels) between the birth position of the $G-$band bright
point and the nearest penumbral feature, i.e, PG or a dark penumbral
area or a dark fibril. The distance of 3 pixels is comparable with
the Rayleigh's resolution limit (0\farcs 23) of a diffraction-limited
47.5\,cm telescope at 430\,nm. The birth positions of 776 $G-$band 
bright points, obtained from
the feature tracking, were compared visually (see examples in Fig.~4)
with the locations of penumbral features in the aligned blue images
(Sect.~2). The blue wavelength band was selected because of higher
contrast. In total, we found 191 $G-$band bright points
born in the neighbourhood with the penumbra and then drifting away
from the spot. Most of them, 132 (69\,\%), originate in the vicinity
of dark penumbral areas or fibrils, where the magnetic field is
expected to be stronger and more horizontal than in bright filaments.
Very often, the $G-$band bright points are born on the tips (ends) of
dark penumbral fibrils. Thus, in the vicinity of a decaying sunspot, most of
the G-Band bright points, which are believed to be associated with
thin magnetic flux tubes, appear to be born close to a magnetized plasma.
Only 59 (31\,\%) $G-$band bright points appear near
bright penumbral features (PGs or bright filaments) or close to
diffuse regions in the penumbra, which are difficult to classify
as ``dark" or ``bright".
Fig.\ 4 represents blue continuum images showing the birth positions of
six $G-$band bright points. Since they are not $G-$band images the
bright points are not evident in the figure and, consequently, white
circles --with the radius of 0\farcs 25-- are employed to indicate their
initial positions or equivalently to outline their neighbourhood.
Black lines represent trajectories of the bright points' motions.
Four of these $G-$band bright points were born on the tips of dark
penumbral fibrils and two in the vicinity of bright PGs.

\begin{figure}
\centering
\includegraphics[width=8cm]{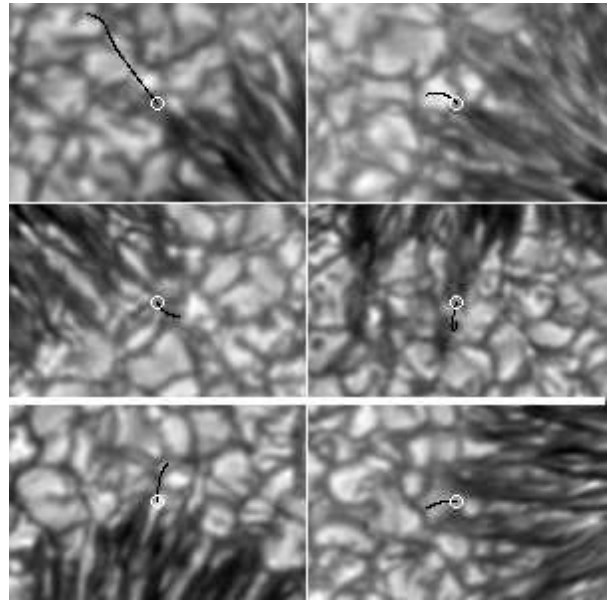}
\caption{Blue continuum images showing the birth positions of four $G-$band 
bright points ({\it top}) born near the tips of dark penumbral fibrils 
and two ({\it bottom}) in the vicinity of bright PGs. $G-$band bright 
points are not evident in the blue continuum so that they are represented
by white circles with radii 0\farcs25 marking their initial positions
and by black lines representing their trajectories. The field of view 
in the images is 12\farcs5 $\times$ 8\farcs3.}
\end{figure}

The small bright features escaping from the penumbra and the $G-$band
bright points born close to the P/G boundary are, at first sight,
similar in morphology and dynamics. To check for possible coincidences
between them, we compared, frame by frame, the positions of PGs after
crossing the P/G border with the positions of all $G-$band bright
points. We did not find any case of coincidence in time and position,
so that we can conclude that escaped PGs and $G-$band bright points
are different types of features and that $G-$band bright points that are
located near the P/G boundary do not originate from bright 
features which crossed the P/G boundary.

In quiet granulation, $G-$band bright points usually appear
in dark spaces which are compressed by expansion or convergent
motions of neighbouring granules (Muller 1983; Muller et al.\ 1989;
Muller \& Roudier 1992). Inspecting the $G-$band movie, we observe
that in 95~\% of all cases, the formation of $G-$band bright points
in the neighbourhood of the penumbral border also occurs in areas
compressed by convergent horizontal motions caused by outward moving 
penumbral features and expanding granules adjacent to the border.
In quiet granulation, the formation of $G-$band
bright points can be attributed to the concentration of diffuse
magnetic flux, already present in the intergranular space, 
by converging or expanding adjacent granules. A similar
mechanism of concentration of magnetic flux could also take place
in the magnetic penumbral plasma at the P/G border.

\section{Discussion and conclusions}

Two time series of 288 frames each, spanning over two hours
and taken simultaneously in the $G-$band and the blue
continuum were restored for instrumental and atmospheric
degradation using the phase-diversity technique. The excellent
quality of the restored images and the stability with time made it
possible to study relations between small-scale penumbral and
photospheric features near the outer penumbral boundary: PGs,
dark penumbral fibrils, granules, and $G-$band bright points.

It was shown that about 1/3 of outward moving PGs located near the
outer penumbral boundary escape from the penumbra and penetrate into
the surrounding granulation where they continue their outward motion,
either as small bright features, or growing as expanding 
granules. Their velocities are slightly higher than 
average speeds in the outer penumbra and in the sunspot
moat. They do not evolve as $G-$band bright points in the sense that
the PGs escaping from the penumbra never coincide spatially with
the $G-$band bright points that are located near the P/G boundary.

While some PGs, belonging to the less-magnetized component in the
penumbra, convert into presumably non-magnetic or weakly magnetized 
granules, many $G-$band bright points are born close to dark penumbral
fibrils (or directly at their tips), where the magnetic field
is strong and nearly horizontal. This confirms the finding by Title et 
al.\ (1995) and Shine et al.\ (1996), reported by Ryutova et al.\ (1998),
that ``moving magnetic features'' frequently appear along the
continuation of dark filaments.
In addition, our observations show that local convergent
motions, caused by the expansion of granules and outward moving
penumbral features, may play a role in their formation. The
compression of a dark space by the surrounding granules, either
in the photosphere away from sunspots, or at the sunspot border,
should be accompanied by a strong downflow, which may help to
concentrate the flux into a magnetic flux tube, detected as a
moving magnetic feature or a $G-$band bright point. It would be of
much interest to obtain high spatial resolution spectra simultaneously
with $G-$band images, to check whether the formation of $G-$band bright
points is associated with strong downflows in dark filaments at the
P/G border of sunspots.

We have not found in the literature any sunspot penumbra model able 
to explain our observations.

\begin{acknowledgements}
J.A.B. and I.M. are grateful to M.G.L\"ofdahl for providing his original
PPDS source code in ANA and for his advice in the construction of the IDL
code basis of the present work. M.S. and J.A.B. thank the Paul Sabbatier 
University in Toulouse for the stay at the Pic du Midi Observatory. R.M. 
and M.S. thank IAC for hospitality. The Swedish Vacuum Solar Telescope 
is operated on the island of La Palma by the Royal Swedish 
Academy of Sciences at the Spanish Observatorio del 
Roque de los Muchachos of the Instituto de Astrof\'\i sica de Canarias. 
The support provided by R. Kever and G. Hosinsky during 
the observations is gratefuly acknowledged.
Partial support by the Spanish Ministerio de Ciencia y Tecnolog\'\i a and
by FEDER through project AYA2001-1649 is gratefully acknowledged.
A part of this work was done in the framework of the research projects
IAA3003404 and K2043105 of the Academy of Sciences of the Czech Republic
and of the project 205/01/0658 of the Grant Agency of the Czech Republic.
Part of this work has been supported by the {\em Deutsche
Forschungsgemeinschaft} under grant PE 782/4 (AT).
This research is part of the European Solar Magnetism Network (EC
contract HPRN-CT-2002-00313).
\end{acknowledgements}


\begin{thebibliography}{}


\bibitem[]{} 
Alissandrakis, C.E., Dialetis, D., Tsiropoula, G., 1987, A\&A 174, 275
\bibitem[]{} 
Beckers, J.M., Schr\"oter, E.H., 1969, Solar Phys. 4,303
\bibitem[]{}
Bellot Rubio, L., 2003, in 3rd International Workshop on Solar
        Polarization, J. Trujillo Bueno, J. S\'anchez Almeida (eds.),
        ASP Conf. Ser., 307, 302
\bibitem[]{}   
Bonet, J.A., 1999, in Motions in the Solar Atmosphere, A. Hanslmeier, \&
M. Messerotti (Eds.), ASSL Ser. 239, (Dordrecht: Kluwer), 1
\bibitem[]{} 
Brickhouse, N.S., La Bonte, B.J., 1988, Solar Phys. 115, 43
\bibitem[]{} 
Denker, C., 1998, Solar Phys. 180, 81
\bibitem[]{} 
Gonsalves, R.A., Childlaw, R., 1979, in  Applications of Digital Image
Processing III, A.G. Tescher (Ed.). Proc. Soc. Photo-Opt. Instrum. Eng.
        207, 32
\bibitem[]{} 
Gonsalves, R.A., 1982, Opt. Eng., 21, 829
\bibitem[]{} 
Harvey, K., Harvey, J., 1973, Solar Phys. 28, 61
\bibitem[]{}
Hirzberger, J., Bonet, J.A., V\'azquez, M., Hanslmeier, A., 1999,
        ApJ 515, 441
\bibitem[]{}
Kawaguchi, I., 1980, Solar Phys. 65, 207
\bibitem[]{}  
Lee, J.W., 1992, Solar Phys. 139, 267
\bibitem[]{} 
L\"ofdahl, M.G., Scharmer, G.B., 1994, A\&A Suppl. 107, 243
\bibitem[]{} 
L\"ofdahl, M.G., Berger, T.E., Shine, R.S., \& Title, A.M., 1998, ApJ 495, 965
\bibitem[]{}
Mart\'\i nez Pillet, V., 1997, in Advances in the Physics of Sunspots,
        B. Schmieder, J.C. del Toro Iniesta, M. V\'azquez (eds.),
        A.S.P. Conf. Ser. 118, 212
\bibitem[]{}
Meyer, F., Schmidt, H.U., Weiss, N.O., Wilson, P.R., 1974, MNRAS 169, 35
\bibitem[]{} 
Molowny-Horas, R., Yi, Z., 1994, ITA (Oslo) Internal Rep. No. 31
\bibitem[]{} 
Muller, R., 1973, Solar Phys. 29, 55
\bibitem[]{}
Muller, R., 1983, Solar Phys. 85, 113
\bibitem[]{}
Muller, R., Hulot, J.C., Roudier, T., 1989, Solar Phys. 119, 229
\bibitem[]{} 
Muller, R., M\'ena, B., 1987, Solar Phys. 112, 295
\bibitem[]{}
Muller, R., Roudier, T., 1992, Solar Phys. 141, 27
\bibitem[]{} 
Paxman, R.G., Schulz, T.J., Fienup, J.R., 1992, J.Opt.Soc.Am. A9, 7, 1072
\bibitem[]{} 
Paxman, R.G., Seldin, J.H., L\"ofdahl, M.G., Scharmer, G.B., Keller, C.U.,
        1996, ApJ, 466, 1087
\bibitem[]{}
Petrovay, K., Moreno-Insertis, F., 1997, ApJ, 485, 398
\bibitem[]{} 
Ryutova, M., Shine, R.A., Title, A.M., Sakai, J.I., 1998, ApJ 492, 402
\bibitem[]{} 
Scharmer, G.B., Brown, D.S., Petterson, L., Rehn, J., 1985, Appl. Opt.
      24, 2558.
\bibitem[]{}
Scharmer, G.B., Gudiksen, B.V., Kiselman, D., L\"ofdahl, M.G., Rouppe
        van der Voort, L.H.M., 2002, Nature 420, 151
\bibitem[]{} 
Schlichenmaier, R. 2002, Astron. Nachr./AN 323, 303
\bibitem[]{} 
Sheeley, N.R., 1969, Solar Phys. 9, 347
\bibitem[]{} 
Sheeley, N.R., 1972, Solar Phys. 25, 98
\bibitem[]{} 
Sheeley, N.R., Bhatnagar, A., 1971, Solar Phys. 19, 338
\bibitem[]{} 
Shine,R.A., Title,A.M., Frank,Z.A., \& Sharmer, G. 1996, BAAS, 28, 871
\bibitem[]{} 
Simon, G.W., Title, A.M., Topka, K.P., Tarbell, T.D., Shine, R.A., Ferguson, S.H,.
     1988, ApJ 327, 964
\bibitem[]{} 
Skumanich, A., 1992, in: Sunspots: Theory and Observations, J.H.
     Thomas, N.O. Weiss (Eds.) (Dordrecht:Kluwer),121
\bibitem[]{} 
Sobotka, M., Brandt, P.N., Simon, G.W., 1997, A\&A 328, 682
\bibitem[]{} 
Sobotka, M., Brandt, P.N., Simon, G.W., 1999, A\&A 348, 621
\bibitem[]{} 
Sobotka, M., S\"utterlin, P., 2001, A\&A 380, 714
\bibitem[]{} 
Title, A.M., Frank, Z.A., Shine, R.A., Tarbell, T.D., Topka, K.P., Scharmer, G.,
     Schmidt, W., 1993, ApJ 403, 780
\bibitem[]{}      
Title,A.M., Frank,Z.A., Shine,R.A., Tarbell,T.D., \& Simon,G.W.
     1995, BAAS, 27, 978.
\bibitem[]{} 
Tonjes, K., W\"ohl, H., 1982, Solar Phys. 78, 63
\bibitem[]{} 
Vrabec, D., 1971, in IAU Symp. 43, Solar Magnetic Fields, R.Howard (Ed.)
     (Dordrecht:Reidel), 329
\bibitem[]{} 
Vrabec, D., 1974, in IAU Symp. 56, Chromospheric Fine Structure, 
     G.Athay (Ed.) (Dordrecht:Reidel), 201
\bibitem[]{} 
Wang, H., Zirin, H., 1992, Solar Phys. 140, 41
\bibitem[]{} 
Westendorp Plaza, C., del Toro Iniesta, J.C., Ruiz Cobo, B., Mart\'\i nez
     Pillet, V., Lites, B.W., Skumanich, A., 2001, ApJ 547, 1130
\bibitem[]{} 
Zirin, H., Wang, H., 1989, Solar Phys. 119, 245



\end{thebibliography}
\end{document}